\documentclass[journal=cmatex,manuscript=article,layout=twocolumn]{achemso}
\usepackage{amsmath,graphics,epsfig,color,verbatim,gensymb}
\usepackage[T1]{fontenc}
\usepackage[utf8]{inputenc}
\usepackage{amssymb}
\DeclareUnicodeCharacter{0146}{\c{n}}

\title{Electronic Structure and Phase Stability of Yb-filled CoSb$_3$ Skutterudite Thermoelectrics from First Principles}
\author{Eric B. Isaacs}
\author{Chris Wolverton}
\email{c-wolverton@northwestern.edu}
\affiliation{Department of Materials Science and Engineering, Northwestern University, Evanston, Illinois 60208, USA}

\begin{document}

\begin{abstract}
Filling the large voids in the crystal structure of the skutterudite
CoSb$_3$ with rattler atoms $R$ provides an avenue for both increasing
carrier concentration and disrupting lattice heat transport, leading
to impressive thermoelectric performance. While the influence of $R$
on the lattice dynamics of skutterudite materials has been well
studied, the phase stability of $R$-filled skutterudite materials and
the influence of the presence and ordering of $R$ on the electronic
structure remain unclear. Here, focusing on the Yb-filled skutterudite
Yb$_x$Co$_4$Sb$_{12}$, we employ first-principles methods to compute
the phase stability and electronic structure. Yb-filled CoSb$_3$
exhibits (1) a mild tendency for phase separation into Yb-rich and
Yb-poor regions and (2) a strong tendency for chemical decomposition
into Co--Sb and Yb--Sb binaries (i.e., CoSb$_3$, CoSb$_2$, and
YbSb$_2$). We find that, at reasonable synthesis temperatures,
configurational entropy stabilizes single-phase solid solutions with
limited Yb solubility, in agreement with experiments. Filling CoSb$_3$
with Yb increases the band gap, enhances the carrier effective masses,
and generates new low-energy ``emergent'' conduction band minima,
which is distinct from the traditional band convergence picture of
aligning the energies of existing band extrema. The explicit presence
of $R$ is necessary to achieve the emergent conduction band minima,
though the rattler ordering does not strongly influence the electronic
structure. The emergent conduction bands are spatially localized in
the Yb-rich regions, unlike the delocalized electronic states at the
Brillouin zone center that form the unfilled skutterudite band edges.
\end{abstract}

%% \date{\today}
%% \maketitle

\section{Introduction}\label{sec:intro}

In thermoelectric heat-to-electricity conversion, the figure of merit
is $ZT = \sigma S^2 T/\kappa$, where $\sigma$ is the electrical
conductivity, $S$ is the thermopower, $\kappa$ is the thermal
conductivity, and $T$ is the temperature. Therefore, efficient
thermoelectric materials must exhibit a rare combination of electronic
and thermal transport properties: large $\sigma$, large $S$, and small
$\kappa$. In order to (1) understand the ability of existing
thermoelectric materials, typically heavily doped semiconductors, to
satisfy this set of rare physical properties and (2) design improved
thermoelectric materials, a detailed understanding of the electronic
structure, lattice dynamics, and phase stability is critically
important.

\begin{figure}[htbp]
  \begin{center}
    \includegraphics[width=1.0\linewidth]{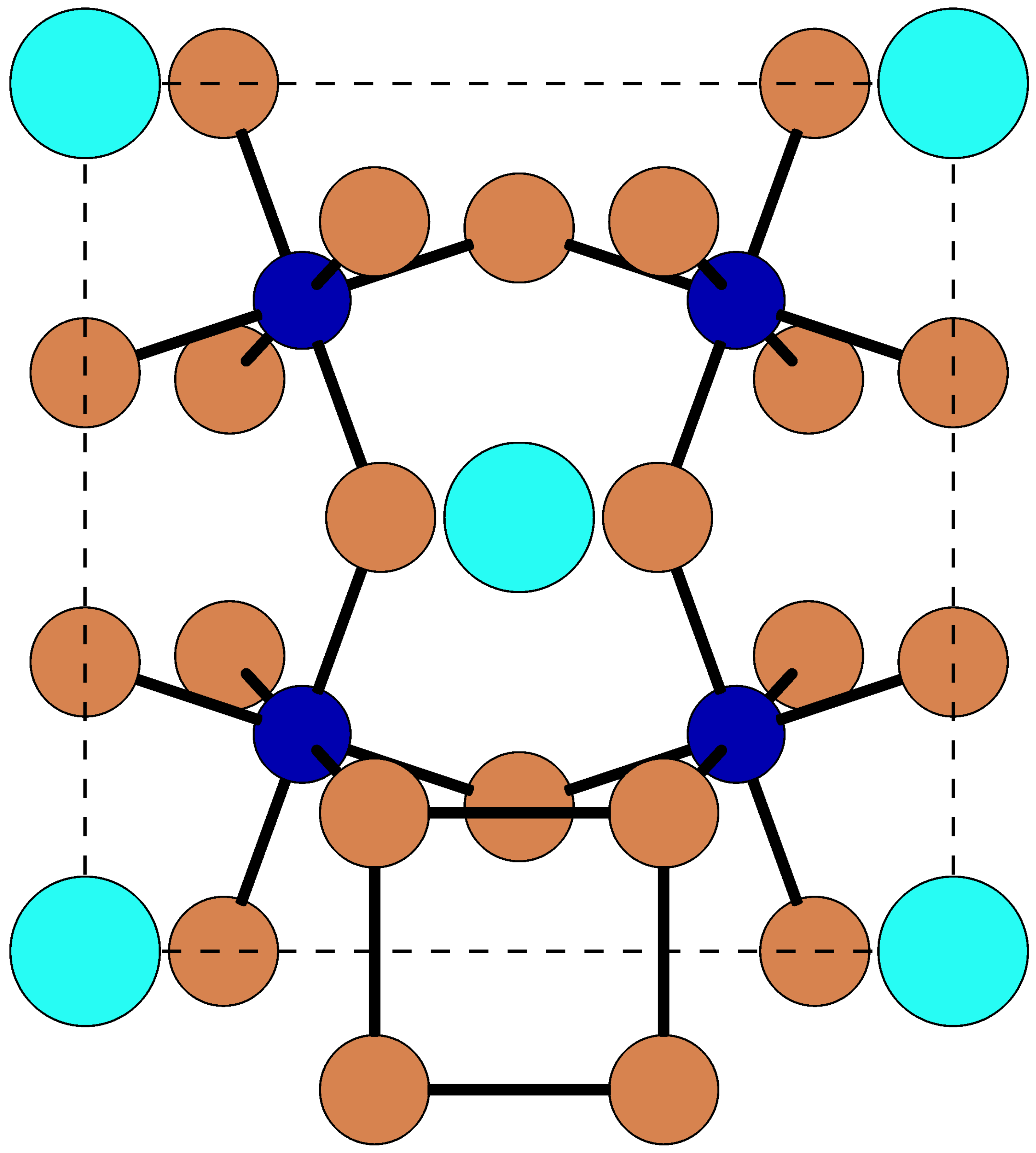}
  \end{center}
  \caption{Skutterudite crystal structure. The conventional unit cell
    is shown with the black dashed lines and cyan, blue, and brown
    circles indicate Yb/vacancy, Co, and Sb, respectively. We note
    that there is an alternative crystal structure description in
    terms of Sb$_4$ rings (sometimes called squares or
    rectangles);\cite{schmidt_structure_1987,lefebvre-devos_bonding_2001}
    one such ring is shown.}
  \label{fig:crystal_structure}
\end{figure}

One famous class of thermoelectric materials is the skutterudite
CoSb$_3$, a covalent semiconductor satisfying the 18-electron
rule.\cite{langmuir_types_1921} CoSb$_3$, whose skutterudite crystal
structure is shown in Fig. \ref{fig:crystal_structure}, can be
considered a perovskite (ABX$_3$, with an empty A-site) with
substantial distortions of the CoSb$_6$ octahedra that create large
voids.\cite{schmidt_structure_1987} CoSb$_3$ has a body-centered-cubic
(bcc) lattice with 16 atoms in the primitive unit cell and a space
group of $Im\bar{3}$. CoSb$_3$-based thermoelectric materials exhibit
favorable electronic transport properties, as the highly covalent
bonding leads to large electronic mobility $\mu$ and $\sigma$ (but
also increasing $\kappa_e$, the electronic contribution to
$\kappa$).\cite{snyder_complex_2008} In addition, the presence of a
high-degeneracy conduction band minimum close in energy to the
conduction band minima at the Brillouin zone center has been invoked
to rationalize the large $S$, via the concept of band
convergence.\cite{tang_convergence_2015,hanus_chemical_2017}

Perhaps the most distinguishing feature of skutterudite materials is
their ability to host ``rattler'' atoms $R$ (such as alkali, alkaline
earth, actinide, rare earth, and halogen elements) in the large
crystallographic voids,\cite{rogl_skutterudites_2017} which serves a
dual purpose with respect to thermoelectricity. First, it enhances the
power factor ($\sigma S^2$) via electronic
doping.\cite{nolas_high_2000} Secondly, it drastically reduces
$\kappa_L$, the lattice component of $\kappa$.\cite{sales_filled_1996}
Loosely bonded to the rest of the solid, the rattler atoms are
believed to disrupt phonon transport via ``rattling'' in the voids
(hence the
name).\cite{slack_properties_1994,nolas_effect_1996,nolas_semiconducting_1998,toberer_phonon_2011}

While the influence of rattlers on the lattice dynamical properties of
skutterudite materials has been much
studied,\cite{feldman_lattice_2000,hermann_einstein_2003,feldman_lattice_2003,viennois_experimental_2005,long_strongly_2005,nolas_assessing_2006,feldman_lattice_2006,wille_antimony_2007,koza_breakdown_2008,wang_resonant_2009,dimitrov_einstein_2010,bernstein_calculations_2010,koza_vibrational_2010,huang_filler-reduced_2010,mochel_lattice_2011,wee_frequency_2012,zebarjadi_role_2012,li_thermal_2014,feldman_lattice-dynamical_2014,koza_vibrational_2014,li_ultralow_2015,koza_low-energy_2015,sergueev_quenching_2015,ren_filling-fraction_2017,fu_collective-goldstone-mode-induced_2018}
the phase stability and electronic properties have received far less
attention. In particular, the thermodynamic stability of filled
skutterudite materials, the ordering tendencies of the rattlers, and
the precise influence of the rattlers on the electronic states are all
unclear. Therefore, in this work, we present a detailed study of the
phase stability and electronic structure of filled skutterudite
CoSb$_3$ using first-principles calculations. We focus on Yb rattlers
since Yb-filled skutterudite CoSb$_3$ exhibits some of the most
promising thermoelectric properties, e.g., $ZT$ approaching 1.5, and
has been subject to considerable experimental
investigation.\cite{rull-bravo_skutterudites_2015,rogl_skutterudites_2017}

We find that the Yb-filled skutterudite exhibits a tendency to phase
separate into Yb-rich and Yb-poor regions, though the energetic
lowering (compared to the completely empty and filled endmembers) is
only on the order of 10 meV per Yb/void site. Due to the small
magnitude of the formation energy, configurational entropy will likely
win this energetic battle, consistent with the single-phase solid
solutions typically found in experiment. The Yb-filled skutterudite is
in a three-phase region of the thermodynamic convex hull, with a
substantial thermodynamic driving force for chemical decomposition
into binaries. We find that this chemical decomposition tendency
limits the Yb solubility, in agreement with experiments. Filling the
CoSb$_3$ skutterudite with Yb opens the electronic band gap, increases
the carrier effective masses, and leads to the emergence of several
new conduction band minima. The explicit presence, though not the
ordering, of the rattlers is responsible for the new conduction band
minima, which are not present in the unfilled skutterudite, as would
be the case for the traditional band convergence picture. The emergent
conduction bands exhibit distinct character with spatial localization
in the Yb-rich regions, as compared to the delocalized electronic
states at the Brillouin zone center.

%% The lowest-energy
%% ordered phases (still higher in energy than phase separation) are
%% (110) and (100) superlattices.

 %% The large Seebeck coefficient in skutterudite has been
 %%  attributed to the presence of a second low-energy conduction band
 %%  with high
 %%  degeneracy.\cite{tang_convergence_2015,hanus_chemical_2017}

%% which serves to a massive
%% reduction in the lattice component of the thermal conductivity
%% ($\kappa_L$) in addition to introducing charge carriers.

\section{Computational Methodology}\label{sec:compdetails}

As can be observed in Fig. \ref{fig:crystal_structure}, CoSb$_3$
exhibits a significant octahedral distortion ($a^+a^+a^+$ in Glazer
notation\cite{glazer_classification_1972}) with respect to the ideal
perovskite structure. While $\angle$Co--Sb--Co is 180\degree\ for
perovskite, it is 127\degree\ in CoSb$_3$. Similarly,
$\angle$Sb--Co--Sb is 85--95\degree\ instead of the ideal 90\degree.
The octahedral distortion in the skutterudite crystal structure yields
1 void per 4 Co atoms. Therefore, the general stoichiometry for a
filled CoSb$_3$ skutterudite is $R_x$Co$_4$Sb$_{12}$ with $0<x<1$.
Using $\square$ to explicitly indicate an empty void, the formula
becomes $\square_{1-x}R_x$Co$_4$Sb$_{12}$. We note that the upper
limit of $x$, i.e., the filling fraction limit (FFL), is lower than 1
in
practice,\cite{yang_solubility_2009,tang_temperature_2015,wang_high-performance_2016,ryll_structure_2018}
but we consider the full crystallographic range of $0\le x\le 1$.

%% CoSb$_3$ consists of corner-sharing CoSb$_6$ octahedra.

Plane-wave density functional theory
(DFT)\cite{hohenberg_inhomogeneous_1964,kohn_self-consistent_1965}
calculations are performed using
\textsc{vasp}\cite{kresse_efficiency_1996} w/ the generalized gradient
approximation of Perdew, Burke, and
Ernzerhof\cite{perdew_generalized_1996} using Co, Sb, and Yb\_2
($5p^66s^2$ valence) projector augmented wave (PAW)
potentials.\cite{blochl_projector_1994,kresse_ultrasoft_1999} We use a
500 eV kinetic energy cutoff, $\Gamma$-centered $k$-point grids of
density $\ge$ 500 $k$-points/\AA$^{-3}$, 0.1 eV 1st-order
Methfessel-Paxton smearing\cite{methfessel_high-precision_1989} for
structural relaxations, and the tetrahedron method with Bl\"{o}chl
corrections\cite{blochl_improved_1994} for static runs. The energy and
ionic forces are converged to 10$^{-6}$ eV energy and 10$^{-2}$
eV/\AA, respectively. Given the reaction energy for Yb + 2CoSb$_3$
$\rightarrow$ YbSb$_2$ + 2CoSb$_2$ changes by only 20 meV/Yb (1.3\%)
via the inclusion of the $4f$ states in the Yb PAW potential, we
expect the absence of such states will not significantly affect our
results.

The convex hull is constructed from the Open Quantum Materials
Database (OQMD),\cite{saal_materials_2013,kirklin_open_2015} a
database of electronic structure calculations based on DFT, which
contains 49 binary and 5 ternary phases in the Yb--Co--Sb space (as of
June 2018). A cluster expansion
(CE)\cite{fontaine_cluster_1994,zunger_first-principles_1994} is
employed to describe the energetics of different configurations of Yb
and $\square$ on the bcc sublattice of voids in the skutterudite
structure. The optimal CE in
\textsc{atat}\cite{van_de_walle_alloy_2002} contains null, point, and
pair (out to 6th nearest neighbor) clusters. Disordered structures are
modeled with special quasirandom structures
(SQS)\cite{zunger_special_1990} with 8 Yb/$\square$ sites for $x$ of
1/4, 1/2, and 3/4.\cite{jiang_first-principles_2004} To generate an
analytical representation of the $x$-dependent energetics of SQS, we
fit the formation energies of the SQS to a Redlich-Kister polynomials
of order 1 (subregular solution model), as discussed in Refs.
\citenum{doak_coherent_2012} and \citenum{doak_vibrational_2015}. Band
structure unfolding based on the CoSb$_3$ lattice parameter is
performed using
\textsc{bandup}.\cite{medeiros_effects_2014,medeiros_unfolding_2015}

% 15 Yb-Co, 16 Co-Sb, 18 Yb-Sb, checked numbers in June 2018 

\section{Results and Discussion}\label{sec:results}

\subsection{Phase stability}

\begin{figure*}[htbp]
  \begin{center}
    \includegraphics[width=1.0\linewidth]{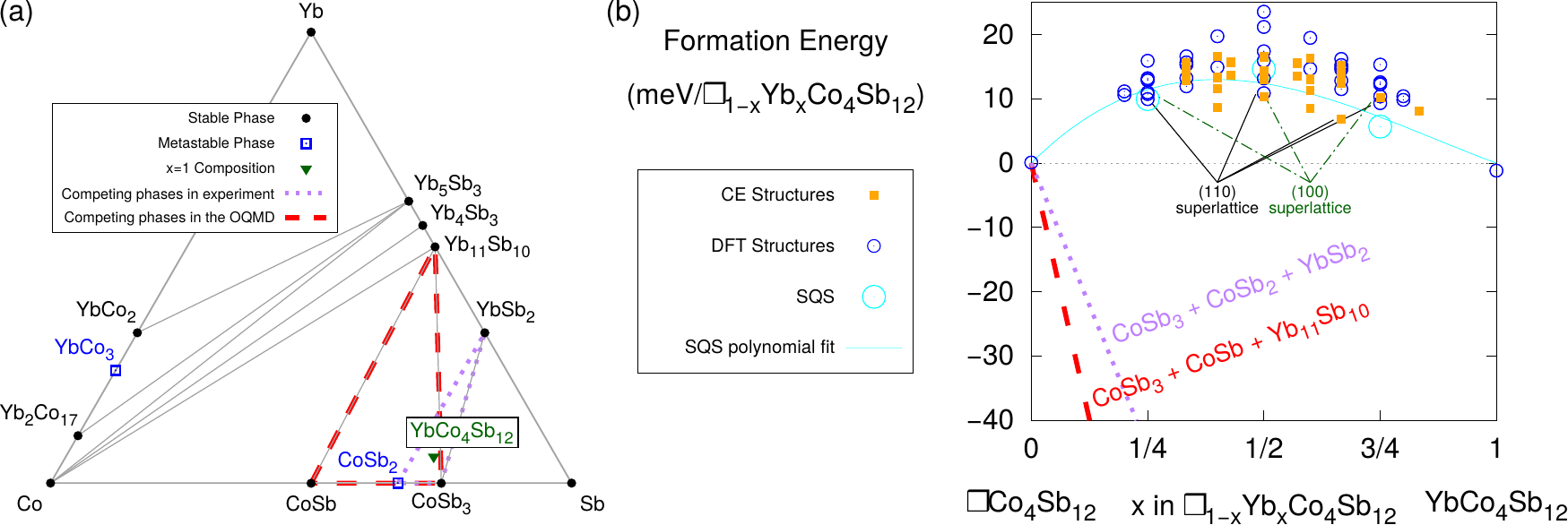}
  \end{center}
  \caption{(a) Yb--Co--Sb ternary convex hull from the OQMD. In
    addition to the stable phases (black filled circles connected by
    grey tie lines), we indicate two metastable phases (above the hull
    by 7 meV/atom for YbCo$_3$, 4 meV/atom for CoSb$_2$) with open
    blue squares. Yb$_x$Co$_4$Sb$_{12}$ corresponds to the line (not
    drawn) between the YbCo$_4$Sb$_{12}$ composition (filled green
    triangle) and CoSb$_3$. The vertices of the red dashed triangle
    indicates the decomposition products for the lowest-energy
    decomposition reaction for Yb$_x$Co$_4$Sb$_{12}$; those of the
    smaller purple dotted triangle correspond to an alternate
    decomposition reaction discussed in the text. (b) Cluster
    expansion formation energy as a function of Yb concentration for
    structures used to fit the cluster expansion (open blue circles)
    and those predicted by the cluster expansion (solid orange
    squares). The DFT-computed formation energy for SQS (large open
    cyan circles) and a polynomial fit (cyan line) are also shown. In
    panel (b), the thick lines in the region of negative formation
    energy correspond to the formation energy of the two decomposition
    reactions indicated in panel (a).}
  \label{fig:phase_stability}
\end{figure*}

Figure \ref{fig:phase_stability}(a), which contains the Yb--Co--Sb
ternary convex hull based on the OQMD, shows that
Yb$_x$Co$_4$Sb$_{12}$ is in a 3-phase region of the convex hull
bounded by CoSb$_3$, CoSb, and Yb$_{11}$Sb$_{10}$. In other words,
Yb$_x$Co$_4$Sb$_{12}$ $\rightarrow (4-5x/11)$ CoSb$_3 + (5x/11)$CoSb
$+(x/11)$Yb$_{11}$Sb$_{10}$ is the lowest-energy decomposition
reaction according to the OQMD. Experiments instead suggest the
competing phases CoSb$_3$, CoSb$_2$, and
YbSb$_2$,\cite{dilley_thermoelectric_2000,tang_temperature_2015} so we
focus on the corresponding decomposition reaction. Although CoSb$_2$
is metastable, it is above the convex hull by only 4 meV/atom. Given
this small energy, it is conceivable that vibrational entropy (not
computed in this work) might stabilize this phase. Artificially
lowering the energy of CoSb$_2$ alone is not sufficient to make
Yb$_x$Co$_4$Sb$_{12}$ $\rightarrow (4-2x)$ CoSb$_3 + 2x$ CoSb$_2 + x$
YbSb$_2$ the lowest-energy decomposition reaction. However, as
discussed in the Supporting Information, its combination with
stabilizing YbSb$_2$ and/or destabilizing Yb$_{11}$Sb$_{10}$ can
achieve this effect. For example, the three-phase equilibrium of
CoSb$_3$, CoSb$_2$, and YbSb$_2$ is achieved by the simultaneous
artificial energy lowering of CoSb$_2$ and YbSb$_2$ by 20 and 23
meV/atom, respectively. Validation of the convex hull from the OQMD is
discussed in the Supporting Information.

%% We note that the treatment of the Yb $4f$
%% electrons as core states is a factor that may influence the computed
%% formation energies.

The formation energies of Yb$_x$Co$_4$Sb$_{12}$, with respect to the
$x=0$ and $x=1$ endmembers, computed via cluster expansion, are shown
in Fig. \ref{fig:phase_stability}(b). The cluster expansion, which is
fit to 40 structures, achieves a leave-one-out cross-validation score
of 1.6 meV per lattice site. Additional details on the cluster
expansion are contained in the Supporting Information. A mild phase
separating tendency is observed, with positive formation energies on
the order of tens of meV per lattice site. Phase separation has also
been predicted in La$_x$Fe$_4$Sb$_{12}$ via coherent potential
approximation calculations (in this case with appreciable energy of
mixing $\sim$ 0.6 eV),\cite{ren_filling-fraction_2017} whereas a
previous cluster expansion for Ba$_x$Co$_4$Sb$_{12}$ found several
stable ordered phases (formation energy no lower than $\sim -90$
meV).\cite{kim_structural_2010} Among the ordered phases, we find
(110) and (100) superlattices are the lowest-energy structures (still
higher in energy than phase separation).

The SQS exhibit DFT-computed formation energies (6--15 meV) similar to
the formation energies of the ordered structures, which is reflective
of the relatively weak interaction between the rattlers. Therefore,
given the small magnitude of the formation energy, we expect
configurational entropy to easily overcome the phase separation
tendency at reasonable synthesis temperatures and enable single-phase
solid solutions of Yb$_x$Co$_4$Sb$_{12}$. For example, the ideal
configurational entropy contribution to the mixing free energy, $-k_BT
[(1-x)\ln(1-x) + x\ln(x)]$, where $k_B$ is the Boltzmann constant, is
$\approx-44$ meV for $x=1/4$ at 900 K, which is significantly larger
in magnitude than the mixing energy ($\approx 10$ meV).

The thermodynamic driving force for chemical decomposition into
binaries, on the other hand, is much stronger than the phase
separation tendency. As shown by the steep negatively sloped lines in
Fig. \ref{fig:phase_stability}(b), the formation energies for chemical
decomposition are significantly larger in magnitude than the mixing
energy. For example, the formation energy for the appropriate linear
combination of CoSb$_3$, CoSb$_2$, and YbSb$_2$ [dotted purple line in
  Fig. \ref{fig:phase_stability}(b)] for $x=1/4$ is $\approx-44$ meV.
Therefore, the chemical decomposition tendency serves to limit the
solubility of Yb in CoSb$_3$, as has been suggested by previous
theoretical
works.\cite{shi_filling_2005,mei_filling_2006,shi_theoretical_2007,shi_thermodynamic_2008}

In experiments, a phase of Yb$_x$Co$_4$Sb$_{12}$ with a maximum in Yb
content is typically achieved without evidence of Yb ordering or
separation into Yb-rich and Yb-poor phases; this phase often coexists
with the binary impurity phases CoSb$_2$ and
YbSb$_2$.\cite{bauer_physical_2000,li_nanostructured_2009,liu_thermoelectric_2012,dahal_thermoelectric_2014,tang_temperature_2015,ryll_structure_2018}
This behavior is qualitatively consistent with our computational
findings. We note that samples whose preparation involves ball milling
may exhibit a inhomogeneous Yb distribution due to non-equilibrium
effects, however.\cite{wenjie_paper}

%% (typical in experimental synthesis\cite{tang_temperature_2015})

\begin{figure}[htbp]
  \begin{center}
    \includegraphics[width=1.0\linewidth]{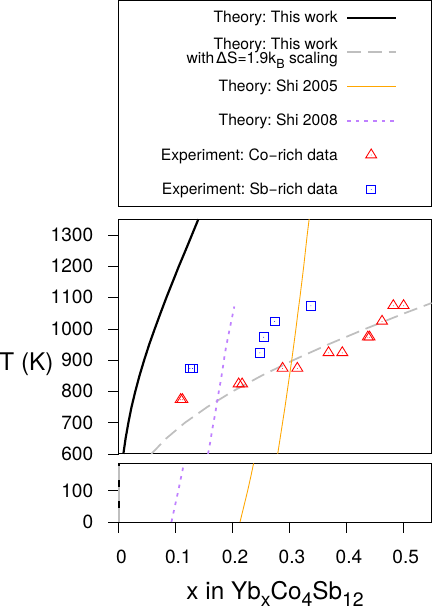}
  \end{center}
  \caption{Temperature-dependent solubility (filling fraction limit)
    of Yb$_x$Co$_4$Sb$_{12}$. The computed solvus from this work is
    shown without (thick solid black line) and with (dashed grey line)
    the scaling factor $e^{\Delta S/k_B}$ for a vibrational entropy
    change $\Delta S = 1.9$ $k_B$, as discussed in the main text. Past
    theoretical solubility predictions from Refs.
    \citenum{shi_filling_2005} (thin solid orange line, labeled Shi
    2005) and \citenum{shi_thermodynamic_2008} (thin dashed purple
    line, labeled Shi 2008) are included for comparison. Experimental
    data for the Co-rich (red open triangles) and Sb-rich (blue open
    squares) region of the experimental phase diagram are taken from
    Fig. 4 of Ref. \citenum{tang_temperature_2015}.}
  \label{fig:solubility}
\end{figure}

Although it has received considerable attention in the literature, the
solubility (FFL) of Yb in CoSb$_3$ remains controversial, with values
reported ranging from 0.2 to
0.7.\cite{dahal_thermoelectric_2014,shi_filling_2005,yang_solubility_2009,he_dynamics_2014,tang_temperature_2015,wang_high-performance_2016}
Here, we address this important issue by computing the solubility from
our first-principles calculations and comparing directly to past
theoretical calculations and recent experiments. In our work, we take
a subregular solid solution model (corresponding to the polynomial fit
to SQS energetics discussed above) and quantitatively incorporate
ideal configurational entropy; vibrational entropy is discussed below.
To compute the solubility, we employ the common tangent construction
with respect to the binary decomposition products observed in
experiment (CoSb$_2$ and YbSb$_2$). Further details on the evaluation
of the solubility from our calculations, as well as details on the
comparison data from past theoretical studies, are included in the
Supporting Information.

Figure \ref{fig:solubility} illustrates our computed (thick solid
black line) temperature-dependent solubility curve (solvus) for Yb in
CoSb$_3$ in comparison with past theory and experiments. The previous
theoretical calculations of Refs. \citenum{shi_filling_2005} (thin
solid orange line) and \citenum{shi_thermodynamic_2008} (thin dashed
purple line) found relatively small temperature dependence. In
addition, the similar previous theoretical work of Mei \textit{et al.}
reported a single FFL value (0.30) rather than temperature-dependent
results.\cite{mei_filling_2006} The lack of strong temperature
dependence in the past theoretical works is in disagreement with the
experimental results of Tang \textit{et al.}, who observed the
measured solubility can vary by as much as a factor of five as a
function of annealing temperature.\cite{tang_temperature_2015}
Importantly, the past theory works also found finite solubility for $T
\rightarrow 0$. Since entropy must vanish for $T \rightarrow 0$ (third
law of thermodynamics), such a solid solution cannot exist on the
$T=0$ phase diagram.\cite{fedorov_third_2010} In this sense, the past
theoretical predictions violate the third law. They also are
inconsistent with experiments, which suggest FFL approaching 0 for $T
\rightarrow 0$.\cite{tang_temperature_2015} We note that it may be
possible to maintain a finite rattler concentration at low $T$ in
experiment, but only as a result of kinetic effects.

In contrast to the past theoretical works, we correctly find a
vanishing solubility at low temperature, which is consistent with the
third law and agrees with experiment.\cite{tang_temperature_2015}
Figure \ref{fig:solubility} shows a quantitative comparison of our
computed solvus with experimental data from Ref.
\citenum{tang_temperature_2015} for the solubility in the Co-rich (red
open triangles) and Sb-rich (blue open squares) regions of the
experimental phase diagram. We focus on the Co-rich data since it
corresponds to equilibrium of Yb$_x$Co$_4$Sb$_{12}$ with CoSb$_2$ and
YbSb$_2$.\cite{tang_temperature_2015} Our computed values appreciably
underestimate the experimental solubility. As a result, although we
find a larger temperature dependence compared to past theoretical
works (and the correct exponential dependence in the dilute limit),
our computed temperature dependence is still significantly smaller
than experiment.\cite{tang_temperature_2015} This solubility
underestimation has been commonly observed in first-principles
predictions due to the neglect of vibrational
entropy.\cite{anthony_vibrational_1993,ozolins_large_2001,ozolins_effects_2005,pomrehn_entropic_2011}
Given Yb is a rattler, corresponding to low-frequency vibrational
modes, one can expect the inclusion of vibrational entropy to
significantly stabilize the Yb$_x$Co$_4$Sb$_{12}$ solid solution,
enhancing the solubility. Via fitting, we find that a vibrational
formation entropy $\Delta S$ (of the solid solution with respect to
the linear combination of binary decomposition products) of $1.9$
$k_B$, taken to modify the solubility via a $e^{\Delta S/k_B}$
multiplicative factor (dilute limit approximation), is sufficient to
reconcile the theoretical prediction with experiment, as shown in Fig.
\ref{fig:solubility}. Experimental measurements and/or calculations of
the phonon entropy will be important future work towards achieving
quantitative solubility prediction for skutterudite materials. We note
that, beyond vibrational entropy, non-ideal configurational entropy
may also help reconcile the computational results with experiment.

%% We include both the Co-rich and Sb-rich data since it is difficult to
%% know which most closely corresponds to the stoichiometric limit due to
%% effects like Sb sublimation.

\subsection{Endmember electronic band structure}

\begin{figure}[htbp]
  \begin{center}
    \includegraphics[width=1.0\linewidth]{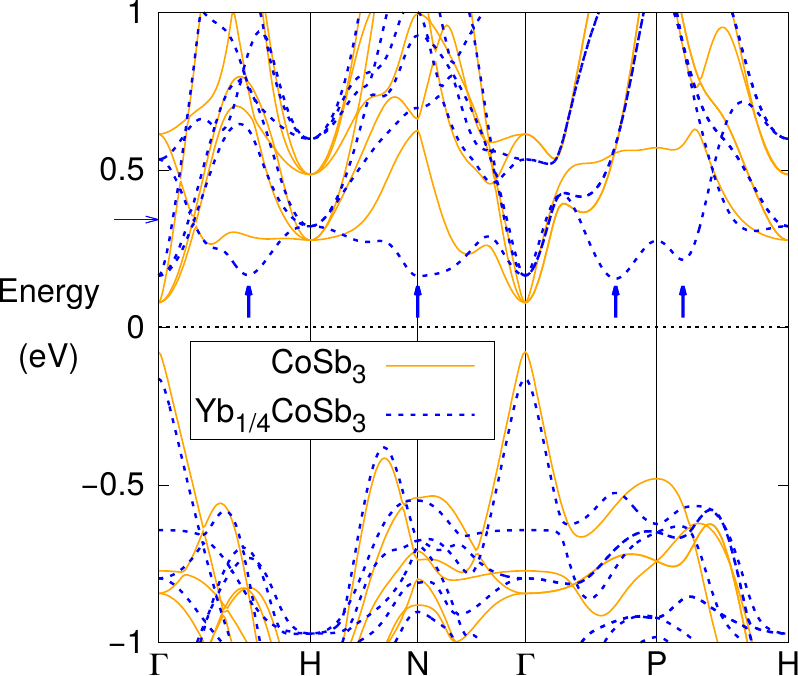}
  \end{center}
  \caption{Electronic band structure of Yb$_x$Co$_4$Sb$_{12}$ for the
    fully-relaxed endmembers ($x=0$ and $x=1$). Energies are plotted
    with respect to the gap midpoint at $\Gamma$, and the Fermi energy
    for $x=1$ is indicated by the horizontal arrow. Emergent
    conduction band minima in $x=1$ are noted by the vertical arrows.}
  \label{fig:endmember_bands}
\end{figure}

The electronic band structure of the endmembers ($x=0$ and $x=1$) is
shown in Fig. \ref{fig:endmember_bands}. Filling the voids with Yb
($x>0$) leads to a metallic state with carriers in the conduction
band. For comparison, the zero of energy set to gap midpoint at
$\Gamma.$ Here, both the $x=0$ and $x=1$ structures are fully relaxed,
so $x=1$ has a smaller Brillouin zone volume than that of $x=1$ due to
a larger lattice parameter. We note that we find the same trends
discussed below if we fix to the relaxed $x=0$ lattice parameter.

With Co contributing 9 e$^-$ and Sb$_3$ contributing 9 e$^-$, CoSb$_3$
satisfies the 18 e$^-$ rule\cite{langmuir_types_1921} and forms a
semiconductor with a small experimental band gap on the order of
35--50 meV.\cite{mandrus_electronic_1995,rakoto_valence_1999} We find
CoSb$_3$ is a direct-gap semiconductor, with the singly-degenerate
valence band and triply-degenerate conduction bands located at
$\Gamma$, consistent with previous
calculations.\cite{singh_skutterudite_1994,sofo_electronic_1998,tang_convergence_2015}
The valence bands are primarily a mix of Co $p/d$ and Sb $p$
character, whereas the conduction bands are primarily Co $d$
character. As shown in early electronic structure calculations on
CoSb$_3$,\cite{singh_skutterudite_1994,sofo_electronic_1998} the
valence band and one of the conduction bands exhibit linear dispersion
(as opposed to the usual parabolic behavior) near the band extrema.

Filling the voids of CoSb$_3$ with Yb leads to two major effects.
First of all, it can be seen that adding Yb increases the magnitude of
the band gap. Secondly, the Yb rattlers lead to the emergence of
additional conduction band minima close in energy to the band edge at
$\Gamma$. We observe such bands, which we refer to as ``emergent
bands'' for reasons discussed in the next section, at four locations
along the high-symmetry $k$-path shown in Fig.
\ref{fig:endmember_bands}: (1) between $\Gamma$ and H, (2) at N, (3)
between P and $\Gamma$, and (4) between P and H. We note that another
emergent band minima exists between N and H, not shown in Fig.
\ref{fig:endmember_bands}. For sufficiently large $x$, the conduction
band minimum no longer corresponds to the band at $\Gamma$, i.e., the
direct gap at $\Gamma$ is no longer the smallest gap. The same trends
of band gap opening and new, emergent conduction band minima are also
present for intermediate $x$ values, as discussed below.

Here, we discuss the band gap trends in more detail. Although
semilocal DFT is well known to exhibit errors in band
gaps,\cite{perdew_density_1985} quasiparticle and spin-orbit coupling
corrections have been shown to yield only small changes to the gap
($\sim 0.1$ eV) in this
system,\cite{khan_electronic_2015,khan_comparative_2017} and we expect
the computed trends to be valid even if there are small errors in the
absolute values. The computed band gap of CoSb$_3$, 0.155 eV, is
larger but still comparable to the small experimental band gap on the
order of 35--50 meV.\cite{mandrus_electronic_1995,rakoto_valence_1999}
Fully filling the voids with Yb (corresponding to the $x=1$ structure)
increases the gap to 0.210 eV. The band gap increases further to 0.239
eV if we fix to CoSb$_3$ structural parameters; this indicates the gap
opening is a chemical, not structural, effect. In order to understand
the role of the Yb atoms, we also artificially dope CoSb$_3$ by
increasing the electron chemical potential of CoSb$_3$ and adding
compensating homogeneous background charge to retain charge
neutrality, rather than including Yb atoms. In such artificially doped
CoSb$_3$, we find a substantial band gap of 0.315 eV, which indicates
that gap opening upon doping is not specifically tied to the presence
of Yb as the rattler. Similar behavior was found in a previous study
of Ba-filled skutterudite CoSb$_3$,\cite{wee_effects_2010} which
suggests the effect is largely invariant to the nature of the rattler.
In contrast to the band gap opening behavior, the emergence of new,
low-energy conduction band minima is \textit{not} found for the
artificially-doped $x=1$ case, whose band structure is shown in the
Supporting Information. This indicates that the presence of the
rattler atoms is responsible for the emergent conduction band minima.

  %% \item Counterintuitive since CoSb$_3$ has a smaller lattice,
  %%   so one might expect smaller gap % due to larger bandwidth

Filling the voids with Yb also impacts the carrier effective masses
$m^*$ determined via a quadratic fit of the band structure near the
band extrema. The effective masses become larger (corresponding to
less dispersive bands) for $x=1$ as compared to $x=0$. For example,
for the valence band at $\Gamma$ along the $P$ direction, the
effective mass is 0.06 $m_e$ for $x=0$ as compared to 0.09 $m_e$ for
$x=1$. Along this direction, there are two heavy and one light
conduction band. For $x=0$, the corresponding effective masses are
0.19 $m_e$ and 0.07 $m_e$, appreciably smaller than the 0.21 $m_e$ and
0.11 $m_e$ for $x=1$, respectively. The same qualitative trend is
found for the $\Gamma$ valence and conduction bands along each of the
$k$-space directions along the computed high-symmetry path in the
Brillouin zone, as is discussed in the Supporting Information. The
decrease in carrier mobility ($\sim 1/m^*$) for larger $x$ is
consistent with experiments.\cite{wenjie_paper}

\subsection{Electronic structure of partially-filled skutterudite CoSb$_3$}

\begin{figure}[htbp]
  \begin{center}
    \includegraphics[width=1.0\linewidth]{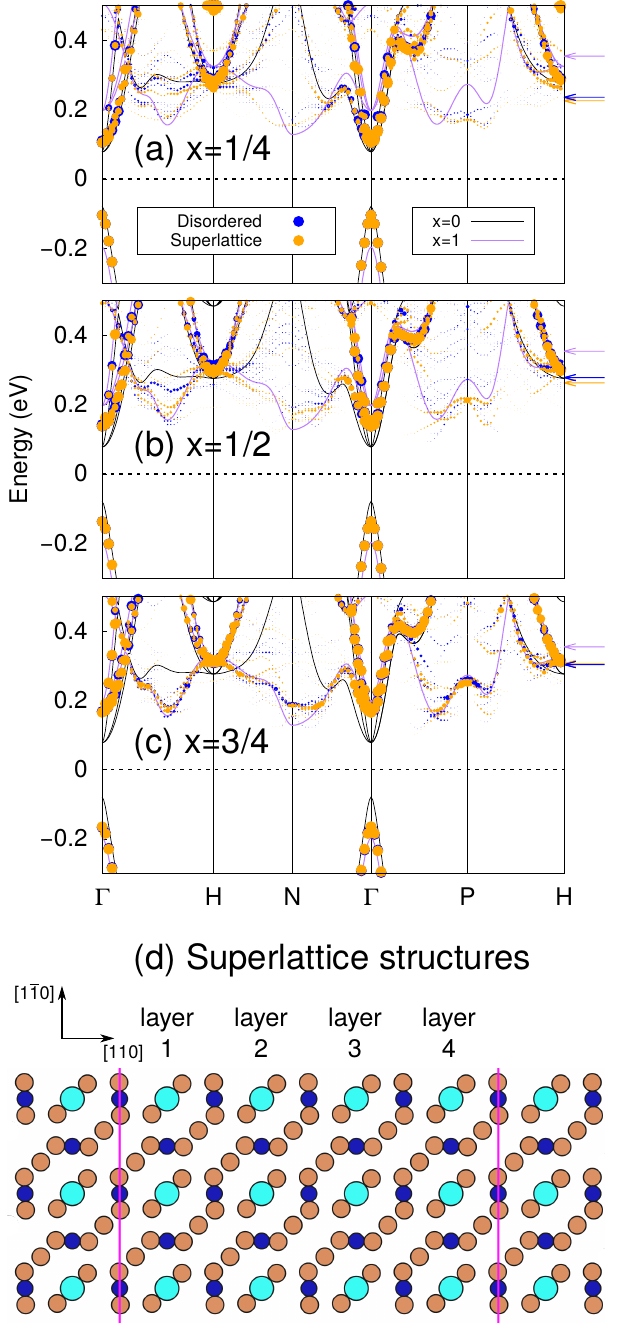}
  \end{center}
  \caption{Electronic band structure of the ordered (110) superlattice
    and disordered structure of Yb$_x$Co$_4$Sb$_{12}$ for (a) $x=1/4$,
    (b) $x=1/2$, and (c) $x=3/4$. Size of the points is proportional
    to the weight in the effective band structure as determined by
    band unfolding. For visual clarity, we only plot points with
    weight greater than 0.02. Fermi energies are indicated by
    horizontal arrows. The endmember band structures are shown as
    solid lines. All structures are fixed to the relaxed $x=0$ lattice
    parameter. The superlattice structures correspond to retaining Yb
    in layer 1 for $x=1/4$, layers 1 and 2 for $x=1/2$, and layers 1,
    2, and 3 for $x=3/4$ as shown in panel (d).}
  \label{fig:ordered_sqs_bands}
\end{figure}

In order to probe the electronic properties of Yb$_x$Co$_4$Sb$_{12}$
with intermediate $x$ ($0<x<1$), we compute the effective band
structures for structures with partial Yb filling, as shown in Fig.
\ref{fig:ordered_sqs_bands}. For ordered structures, we choose the
low-energy structures corresponding to (110) superlattices since such
structures have a relatively small primitive unit cell. In addition,
we note that previous DFT calculations found that (110) is the
lowest-energy surface.\cite{hammerschmidt_low-index_2015} We compare
to disordered structures in order to assess the effect of Yb ordering
on the electronic structure. Both the ordered and disordered
structures show similar effects as the fully-filled skutterudite
material: band gap opening and emergent conduction band minima. This
suggests rattler ordering does not have a dramatic effect on the band
structure, though the presence of the rattler atoms is necessary to
achieve the new conduction band minima (as discussed above). We note
that there are differences between the ordered and disordered
structures in the finer details of the emergent bands.

\begin{figure}[htbp]
  \begin{center}
    \includegraphics[width=1.0\linewidth]{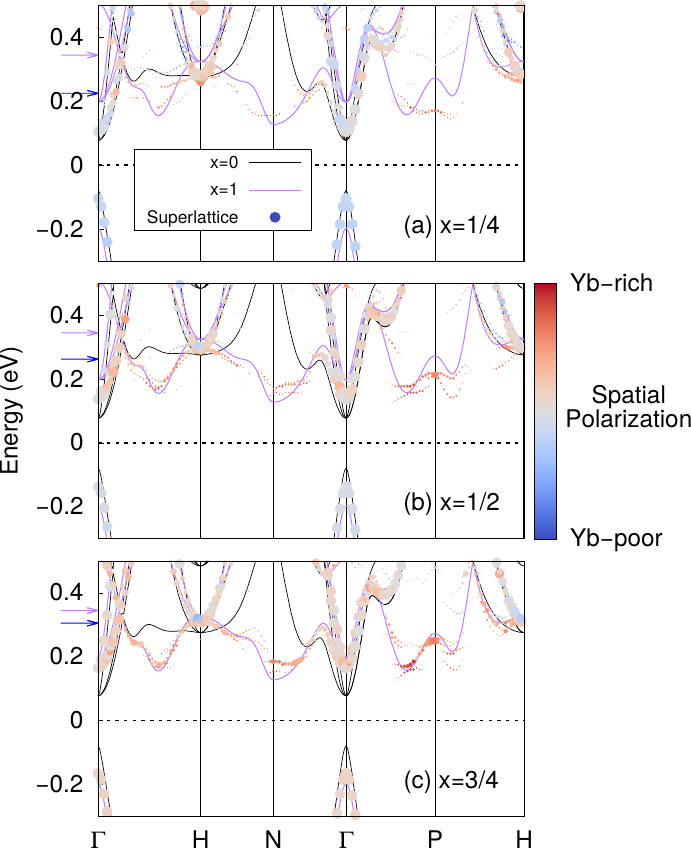}
  \end{center}
  \caption{Electronic band structure of the ordered (110) superlattice
    of Yb$_x$Co$_4$Sb$_{12}$ for (a) $x=1/4$, (b) $x=1/2$, and (c)
    $x=3/4$. Size of the points is proportional to the weight in the
    effective band structure as determined by band unfolding, and the
    color indicates the spatial polarization $\xi$ of the
    wavefunctions (as defined in the main text). We only plot points
    with weight greater than 0.02 (for visual clarity) whose
    $p_{\mathrm{Yb-rich}}$ and $p_{\mathrm{Yb-poor}}$ are both greater
    than 0.025 (to avoid numerical errors in computing $\xi$). Fermi
    energies are indicated by horizontal arrows. The endmember band
    structures are shown as solid lines. All structures are fixed to
    the relaxed $x=0$ lattice parameter.}
  \label{fig:spatial_polarization}
\end{figure}

In order to investigate the nature of the electronic states in the
partially-filled skutterudite material, we compute for the
superlattice structures the projections of the wavefunctions (1) on
all atoms corresponding to the Yb-rich region (which we call
$p_{\mathrm{Yb-rich}}$) and (2) on all atoms corresponding to the
Yb-poor region (which we call $p_{\mathrm{Yb-poor}}$). The layers of
atoms at the interfaces between these regions are not included in
either of these projections. For example, for $x=1/2$, these interface
atoms include those lying in the purple planes drawn in Fig.
\ref{fig:ordered_sqs_bands}(d) as well as the corresponding atoms
between layers 2 and 3. We define the \textit{spatial polarization},
with respect to the Yb-rich and Yb-poor regions, of the electronic
states as $$\xi = \frac{p_{\mathrm{Yb-rich}}/x}{p_{\mathrm{Yb-rich}}/x
  + p_{\mathrm{Yb-poor}}/(1-x)}.$$ The factors of $1/x$ and $1/(1-x)$
are included to normalize for the differing sizes of the Yb-rich and
Yb-poor regions when $x \ne 1/2$. A spatial polarization value of 1/2
indicates the wavefunction exhibits an equal preference for
localization on an atom in the Yb-rich region as that in the Yb-poor
region, whereas a $\xi$ value of 1 (0) indicates a 100\% preference
for localization on an atom in the Yb-rich (Yb-poor) region.

%%  For example, $p_{\mathrm{Yb}} =
%% \sum_{\alpha, l, m} | \langle Y_{lm}^{\alpha}|
%% \psi_{n\vec{k}}\rangle|^2,$ where $\psi_{n\vec{k}}$ is the Kohn-Sham
%% wavefunction for band $n$ and crystal momentum $\vec{k}$ and
%% $Y_{lm}^{\alpha}$ is a spherical harmonic centered on an atom $\alpha$
%% in the Yb-rich region with angular momentum (projection) $l$ ($m$); we
%% sum over all values of $l$ and $m$ in order to provide a measure for
%% the total charge in the region.

The spatial polarization of the superlattice electronic states is
shown via the color of the points in Fig.
\ref{fig:spatial_polarization}, which shows the superlattice band
structure. The emergent conduction bands exhibit values of $\xi$
significantly larger than 1/2, which indicates such states tend to be
localized in the Yb-rich region. Although these states have a strong
preference to localize in the Yb-rich \textit{region}, we note that
the states are not localized on the Yb atoms, which act as cations and
donate their charge. In contrast to the emergent bands, the electronic
states at $\Gamma$ show values of $\xi$ much closer to 1/2. This
indicates that the highly-dispersive bands at $\Gamma$ are much more
spatially delocalized. Therefore, one can think of
Yb$_x$Co$_4$Sb$_{12}$ as containing two distinct types of carriers:
delocalized electrons at $\Gamma$ and electrons more localized in the
Yb-rich regions from the emergent bands.

Finally, we discuss our use of the term ``emergent bands'' and put our
results in the context of the concept of band convergence. The
behavior we find is quite distinct from typical band convergence (such
as in PbTe$_{1-x}$Se$_x$\cite{pei_convergence_2011},
Mg$_2$Si$_{1-x}$Sn$_x$\cite{liu_convergence_2012},
Pb$_{1-x}$Mg$_x$Te\cite{zhao_all-scale_2013}, and
Pb$_{1-x}$Sr$_x$Se\cite{wang_tuning_2014}) in which the energies of
multiple existing band minima converge as a function of some tuning
parameter, such as temperature or doping. In our case, several of the
low-energy minima away from $\Gamma$ are not in general even present
(in any recognizable form) in the unfilled material.

For example, as can be seen in Fig. \ref{fig:endmember_bands} between
$\Gamma$ and P, the lowest-energy conduction band minimum for $x=1$
($\sim 2/3$ of the way from $\Gamma$ to P), which is associated with
the lowest-energy minima along this high-symmetry line in $k$-space
for fractional $x$ in Figs. \ref{fig:ordered_sqs_bands}(a-c),
essentially does not exist for $x=0$. In other words, this band minima
does not appear at all similar to the band from which it appears to
originate, which is the very flat (away from $\Gamma$) $x=0$ band at
energy of $\sim 0.5$ eV in Fig. \ref{fig:endmember_bands}. In this
sense, such bands ``emerge'' rather than ``converge'' with Yb filling
and we describe the new band minima as ``emergent bands'' rather than
``convergent bands.''

The band convergence previously discussed in the literature for
skutterudites\cite{tang_convergence_2015,hanus_chemical_2017,korotaev_influence_2017}
takes a different form than that identified in our work. Since in
these previous works the term was applied to the convergence of the
energy of an existing conduction band minimum between $\Gamma$ and N
for $x=0$ (i.e., the lowest-energy minimum roughly halfway from
$\Gamma$ to N in Fig. \ref{fig:endmember_bands}) to that of the
conduction band minima at $\Gamma$, this corresponds to the typical
use of the band convergence term, as discussed above. Our results
strongly suggest that several other band minima, including those
emergent bands absent in the unfilled material, should also
substantially contribute to the electronic transport. Indeed, as shown
in Fig. \ref{fig:ordered_sqs_bands}(a), even for the lowest filling
value considered of $x=1/4$, these other band minima are significantly
lower in energy than the band minimum previously considered in the
literature.

\section{Conclusions}
Using first-principles calculations, we provide a detailed
understanding of the phase stability and electronic properties of
filled skutterudite CoSb$_3$. The Yb-filled skutterudite
Yb$_x$Co$_4$Sb$_{12}$ exhibits a mild tendency to phase separate into
the Yb-rich and Yb-poor endmembers, as well as a strong tendency for
chemical decomposition into Co--Sb and Yb--Sb binaries. Single-phase
solid solutions with a limited Yb solubility, observed in experiment,
are stabilized by configurational entropy. In addition to enhancing
the band gap and effective masses, the presence of Yb leads to two
distinct types of electronic carriers: (1) new emergent conduction
band minima whose electronic states are localized near the rattler
atoms and (2) the delocalized electronic states at the Brillouin zone
center.

\begin{acknowledgement}
We thank Jeff Snyder (Northwestern) and Wenjie Li (Penn State) for
useful discussions. We acknowledge support from the U.S. Department of
Energy under Contract DE-SC0014520. Computational resources were
provided by the National Energy Research Scientific Computing Center
(U.S. Department of Energy Contract DE-AC02-05CH11231), the Extreme
Science and Engineering Discovery Environment (National Science
Foundation Contract ACI-1548562), and the Quest high performance
computing facility at Northwestern University.
\end{acknowledgement}

\begin{suppinfo}
Additional details on the Yb--Co--Sb convex hull,
Yb$_x$Co$_4$Sb$_{12}$ cluster expansion, polynomial fit to formation
energies of SQS, computation of solvus and comparison to past
solubility predictions, the electronic band structure of artificially
doped CoSb$_3$, and the carrier effective masses for CoSb$_3$ and
YbCo$_4$Sb$_{12}$.
\end{suppinfo}

\bibliography{skutterudites}

%% \begin{tocentry}
%%   \includegraphics[width=8.47cm]{toc.png}
%%   %% \includegraphics[height=4.76cm]{toc.png}
%% \end{tocentry}

\end{document}